\documentclass[final,10pt]{elsarticle}

\usepackage[dvipsnames,usenames]{color}

\usepackage[numbers]{natbib}

\usepackage{array}
\usepackage{amsmath}
\usepackage{amssymb}
\usepackage{url}
\usepackage{multirow}
\usepackage{array}
\usepackage{graphicx}
\usepackage{subfigure}
\usepackage[normalem]{ulem}
\usepackage{subfig}
\usepackage{caption}
\usepackage{float}

\usepackage{listings}
\definecolor{dkgreen}{rgb}{0,0.6,0}
\definecolor{gray}{rgb}{0.5,0.5,0.5}
\definecolor{mauve}{rgb}{0.58,0,0.82}

\usepackage{algorithm}
\usepackage{algpseudocode}
\usepackage{pifont}
\usepackage{marginnote}
\usepackage{color}

\usepackage{amsthm}
\makeatletter
\def\theequation{\arabic{equation}}
\makeatother



\journal{}

\begin{document}
\begin{frontmatter}
  \title{BitCracker: BitLocker meets GPUs}
%\title{BitLocker Dictionary Attack using GPUs}

\author{E.~Agostini$^\dagger$, M.~Bernaschi$^\dagger$}
\address{$^\dagger$National Research Council of Italy (CNR), Rome
  (ITALY)}

\begin{abstract}
BitLocker is a full-disk encryption feature available in recent Windows
versions. It is designed to protect data by providing encryption for
entire volumes and it makes use of a number of different authentication methods.
In this paper we present a solution, named BitCracker, to attempt the
decryption, by means of a dictionary attack, of memory units
encrypted by BitLocker with a user supplied password or the recovery password.
To that purpose, we resort to GPU (Graphics Processing Units) that
are, by now, widely used as general-purpose coprocessors in high performance computing applications.
BitLocker decryption process requires the computation of a very large number
of SHA-256 hashes and also AES, so we propose a very fast solution,
highly tuned for Nvidia GPU, for both of them.
We analyze the performance of our CUDA implementation on several Nvidia GPUs and we carry out a comparison of our SHA-256 hash with
the Hashcat password cracker tool. Finally, we present our OpenCL version, recently
released as a plugin of the John The Ripper tool.

\end{abstract}

%\maketitle
\end{frontmatter}

\pagebreak

\section{Introduction}\label{sec:introduction}

BitLocker is a data protection feature that integrates with the
Windows operating system and addresses the threats of data theft or
exposure from lost, stolen, or inappropriately decommissioned
computers. It offers a number of different authentication methods,
like Trusted Platform Module, Smart Key, Recovery Password, user supplied password.
BitLocker features a pretty complex proprietary architecture but it
also leverages some well-known algorithms, like SHA-256 and AES. It is
possible, and relatively easy (to the purpose, commercial tools are
available \cite{bib:elcomsoft}) to instantly decrypt disks and volumes
protected with BitLocker by using the decryption key extracted from
the main memory (RAM). In addition, it is also possible to decrypt for
offline analysis or instantly mount BitLocker volumes by utilizing the
escrow key (BitLocker Recovery Key) extracted from a user’s Microsoft
Account or retrieved from Active Directory.

If the decryption key can not be retrieved, the only alternative
remains to unlock password-protected disks by attacking the password.
The same, above mentioned, commercial tools offer this as an option
but in a quite generic form ({\em i.e.,} without taking into account
the specific features of BitLocker). Moreover, according to some
comments\footnote{https://blog.elcomsoft.com/2016/07/breaking-bitlocker-encryption-brute-forcing-the-backdoor-part-ii/},
they may be also not fully reliable. The goal of the present paper is
to describe our approach to attack BitLocker password-protected
storage units. We carefully studied available information about
BitLocker architecture and directly inspected several types of units
in order to find out how to minimize the amount of work required to
check a candidate password. The platforms we use for the attack are
based on Nvidia GPUs and we carefully optimized the most computing
intensive parts of the procedure achieving a performance that is, at least,
comparable with that provided by well-know password crackers like
Hashcat \cite{bib:hashcat} for the evaluation of the
SHA-256 digest function. However the main goal of our work is not providing an
alternative to Hashcat as a general framework for dictionary attacks
but to offer the first open-source high performance tool to test the
security of storage units protected by BitLocker using the user password and recovery password
authentication methods. The rest of the paper is organized as follows:
Section \ref{sec:bitlocker} describes BitLocker and in particular the decryption
procedure of the so-called {\em Volume Master Key}; Section \ref{sec:bitcracker}
describes our attack, that we name {\em BitCracker}, to BitLocker focusing
on the optimizations made to improve the performance of the execution of
the SHA-256 algorithm (that is the computational bottleneck of the decryption procedure);
Section \ref{sec:cudaimpl} presents the performance (as number of passwords that
is possible to check {\em per} second) of BitCracker using different variants
of the CUDA \footnote{https://developer.nvidia.com/cuda-zone} architecture showing that there is an improvement of more than a factor 8
moving from the Kepler to the Volta architecture. Section \ref{sec:opencl} presents our OpenCL \footnote{https://www.khronos.org/opencl} implementation
comparing the password rate with results in Section \ref{sec:cudaimpl}.
Section \ref{sec:performance} reports a comparison of the performance, limited to the evaluation of the SHA-256 digest,
of BitCracker with respect to Hashcat. Finally Section \ref{sec:conclusions} summarizes
the results and provides indications for future activities.

\section{BitLocker}\label{sec:bitlocker}

BitLocker (formerly BitLocker Drive Encryption) is a full-disk
encryption feature included in the Ultimate and Enterprise editions
of Windows Vista and Windows 7, the Pro and Enterprise editions of
Windows 8 and Windows 8.1, Windows Server 2008 and Windows 10. It is designed to protect data by providing encryption for entire volumes.

\begin{figure}[h]
\centering
\includegraphics[width=\linewidth]{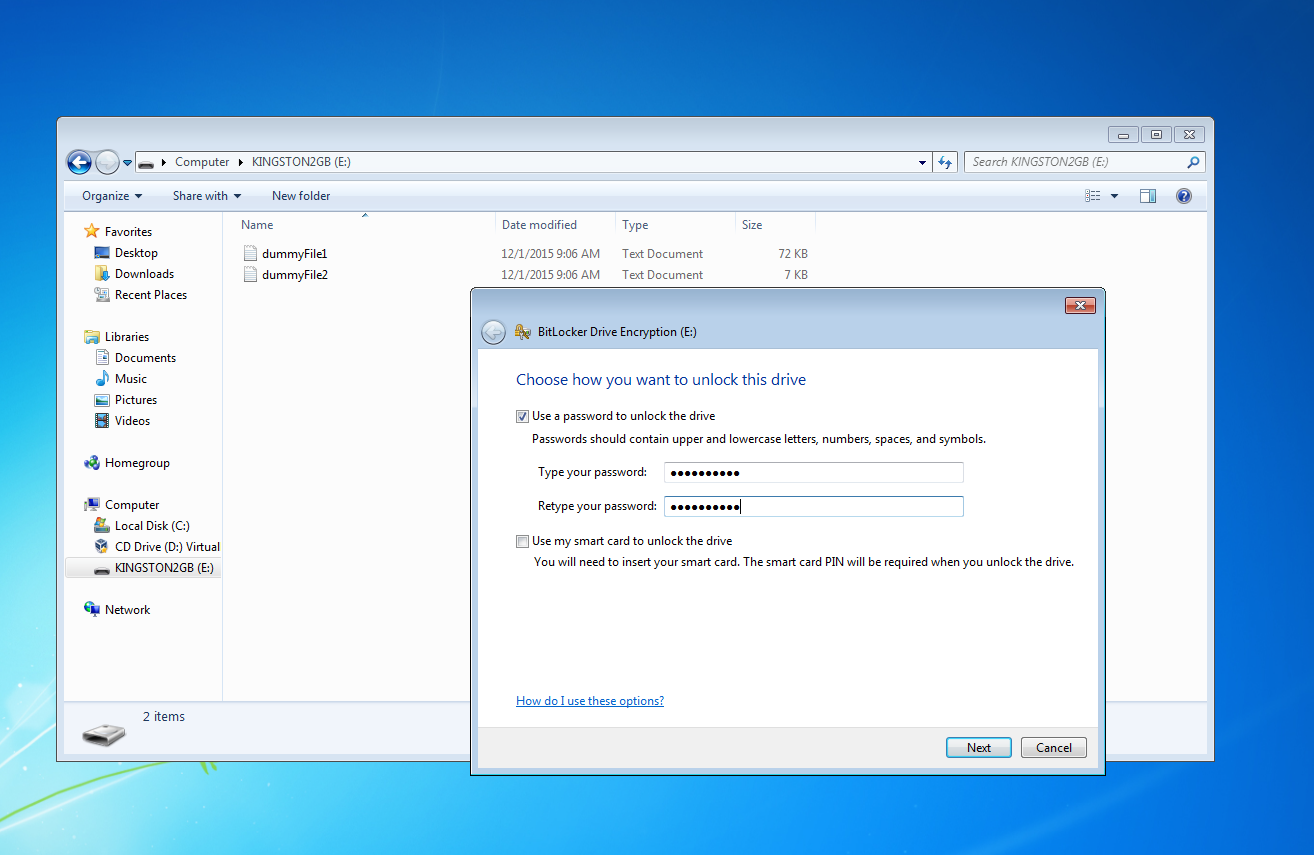}
\caption{BitLocker encryption of an USB pendrive using the password authentication method.\label{fig:screen}}
\end{figure}

BitLocker can encrypt several types of memory units like internal hard disks or external memory devices \footnote{BitLocker To Go feature}(flash memories, external hard disks, etc..)
offering a number of different authentication methods, like Trusted Platform Module, Smart Key, Recovery Key, password, etc..
In this paper we focus on two different authentication modes: the \textit{user password mode}, in which the user, to encrypt or decrypt a memory device, must type a password (as represented in Figure \ref{fig:screen}) and the \textit{recovery password mode}, that is a 48-digit key generated by BitLocker (regardless of the authentication method chosen by the user) when encrypting a memory device\footnote{Microsoft Blog: Recover Password method: https://docs.microsoft.com/en-us/windows/device-security/bitlocker/bitlocker-recovery-guide-plan} . By means of the recovery password the user can access an encrypted device in the event that she/he can't unlock the device normally.

%BitLocker supports several alternatives for the encryption of a whole device, like removable hard disks \footnote{•}, USB devices or internal hard disks ~\cite{bib:bitlockerdriveencryption} like Trusted Platform Module, Smart Key, Recovery Key, password, etc.. In this paper we focus on the password mode, in which the user, to encrypt or decrypt a memory device, must type a password as represented in Figure \ref{fig:bitlocker_enc_dec}.

\iffalse

\begin{figure*}[htp]
  \centering
  \subfigure[BitLocker Encryption]{\includegraphics[scale=0.3]{bitlocker_encrypt.png}}\quad
  \subfigure[BitLocker Decryption]{\includegraphics[scale=0.3]{bitlocker_decrypt.png}}
  \caption{BitLocker Encryption and Decryption procedure of an USB
    pendrive based on a user password.}
  \label{fig:bitlocker_enc_dec}
\end{figure*}

\fi

During the encryption procedure, each sector in the volume is encrypted individually, with a part of the encryption key being derived from the sector number itself. This means that two sectors containing identical unencrypted data will result in different encrypted bytes being written to the disk, making it much harder to attempt to discover keys by creating and encrypting known data.
BitLocker uses a complex hierarchy of keys to encrypt devices. The
sectors themselves are encrypted by using a key called the
\emph{Full-Volume Encryption Key} (FVEK). The FVEK is not used by or
accessible to users and it is, in turn, encrypted with a key called the
\emph{Volume Master Key} (VMK). This level of abstraction gives some
unique benefits, but it makes the process a bit more difficult to
understand. The FVEK is kept as a closely guarded secret because, if
it were compromised, all of the sectors would need to be re-encrypted.
Since that would be a time-consuming operation, it is much better to
avoid it. Instead, the system works with the VMK.
The FVEK (encrypted with the VMK) is stored on the disk itself, as part of the volume metadata and it is never written to disk unencrypted. The VMK is also encrypted with one or more (combination of)
authentication mechanisms as above mentioned; for instance, if the memory device has been encrypted with
the user password method, in the volume metadata there are two encrypted VMKs: the VMK\_U, that is the VMK encrypted with the user password, and the VMK\_R, that is the VMK encrypted with the recovery password.
Both FVEK and VMK are encrypted according to the Counter with CBM-MAC (CCM) mode of AES.

%In Figure \ref{fig:bitlockerscheme}, we show the  BitLocker decryption scheme: depending on the authentication method, there are different algorithms to decrypt the VMK key.

\begin{figure}[h]
\centering
\includegraphics[width=\linewidth]{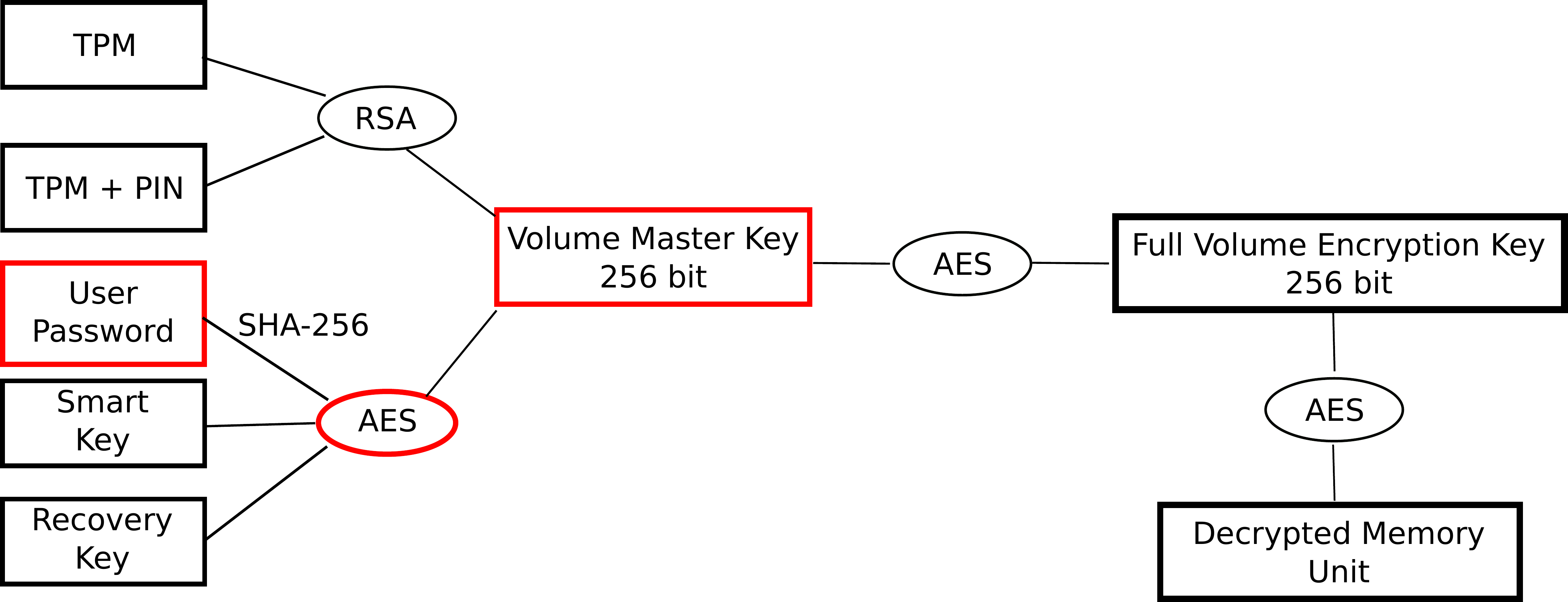}
\caption{BitLocker encryption/decryption scheme\label{fig:bitlockerscheme}}
\end{figure}

During the decryption procedure (Figure \ref{fig:bitlockerscheme}) BitLocker,
depending on the authentication method in use, starts to decrypt the VMK.
Then, if it obtains the right value for the VMK, it decrypts in turn the FVEK and then the entire memory device.

The attack described in the present paper aims at decrypting the
correct VMK key which belongs to an encrypted memory unit through a
dictionary attack to the user password or to the recovery password. That is, if an attacker is
able to find the password to correctly decrypt the VMK key, she/he is able to
decrypt the entire memory unit with that password.

\subsection{User Password VMK Decryption Procedure\label{sec:vmkdecryptionprocedure}}

\begin{figure}[h]
\begin{center}
\includegraphics[width=\linewidth]{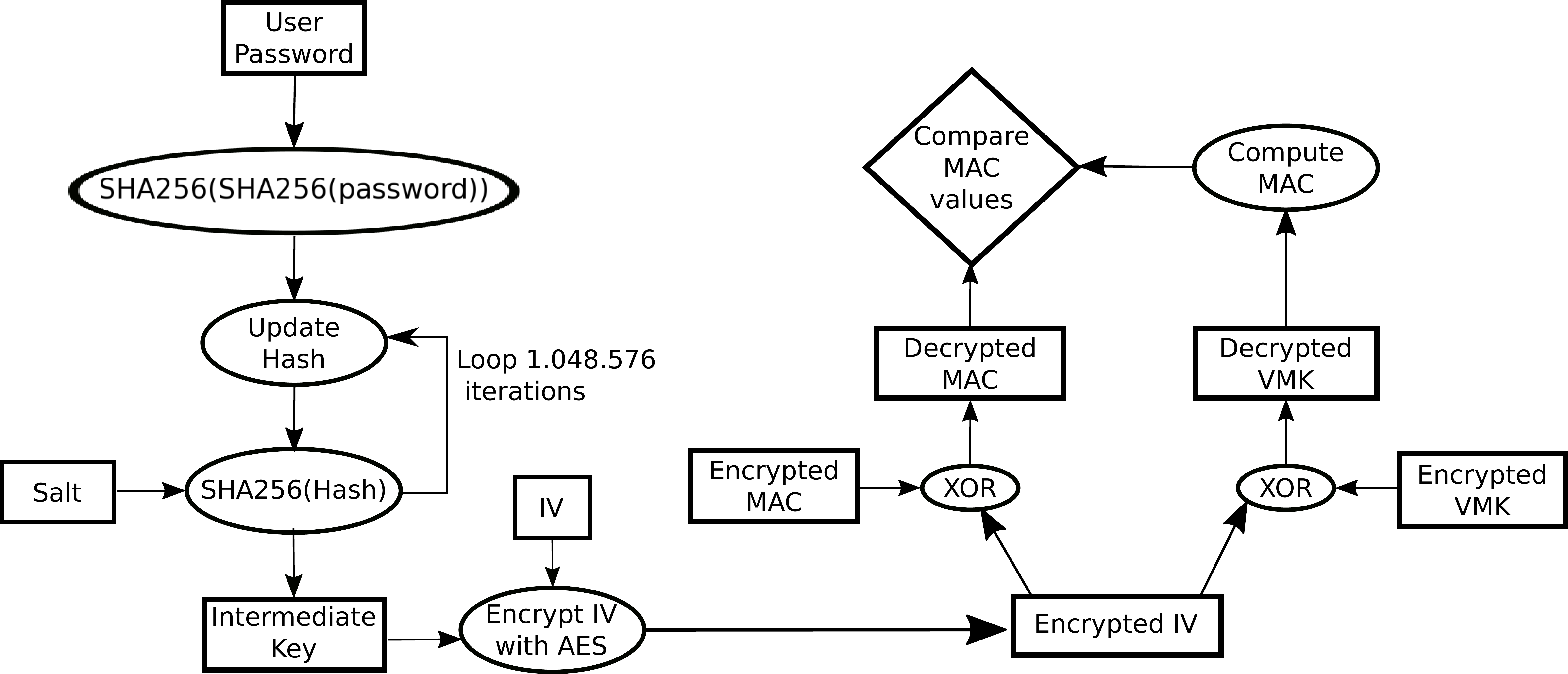}
\caption{VMK decryption procedure\label{fig:vmkdecrypt1}}
\end{center}
\end{figure}

To gain an insight about the workings of our attack, more information are necessary about
the VMK decryption procedure (Figure \ref{fig:vmkdecrypt1}) when the authentication method is a user password
(see also \cite{bib:kumar}  \cite{bib:dislocker} and \cite{bib:libbde}):
\begin{enumerate}
\item the user provides the password;
\item SHA-256 is executed twice on it;
\item there is a loop of 0x100000 iterations, in which SHA-256 is applied to a structure like:
\begin{lstlisting}
typedef struct {
    unsigned char updateHash[32]; //last SHA-256 hash calculated
    unsigned char passwordHash[32]; //hash from step 2
    unsigned char salt[16];
    uint64_t hash_count; // number of hash in loop, incremented
                            // at the end of every iteration
} bitlockerMessage;
\end{lstlisting}
\item this loop produces an intermediate key, used with AES to encrypt
  the Initialization Vector (IV) (derived from a {\em nonce});
\item XOR between encrypted IV and encrypted Message Authentication
  Code (MAC) to obtain the decrypted MAC;
\item XOR between encrypted IV and encrypted VMK to obtain the decrypted VMK;
\item if the MAC, calculated on the decrypted VMK, is equal to the decrypted MAC, the input password and the decrypted VMK are correct;
\end{enumerate}

All the elements required by the decryption procedure (like VMK, MAC, IV, etc..) can be found inside the encrypted volume.
In fact during the encryption, BitLocker stores not only encrypted data but also metadata
that provide information about encryption type, keys position, OS
version, file system version and so on.
Thanks to \cite{bib:libbde},  \cite{bib:dislocker}, \cite{bib:kumar} and \cite{bib:implbitlocker} we understood how to get all of those information
reading the BitLocker Drive Encryption (BDE) encrypted format.

After an initial header, every BDE volume contains 3 (for backup purposes) FVE (Full Volume Encryption) metadata blocks,
each one composed by a block header, a metadata header and an array of metadata entries.
\begin{figure}[h]
\begin{center}
\includegraphics[width=\linewidth]{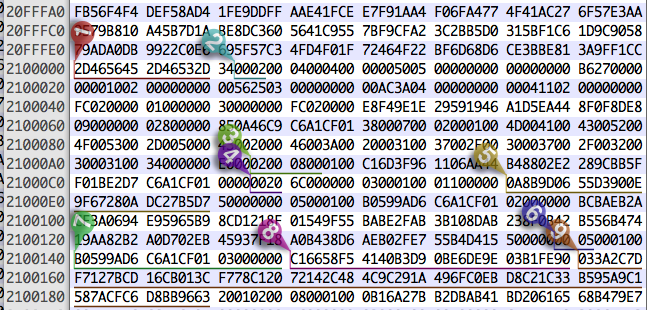}
\caption{FVE metadata block, BitLocker Windows 8.1\label{fig:fvevmk}}
\end{center}
\end{figure}

In Figure \ref{fig:fvevmk} we report an example of FVE block belonging to a memory unit encrypted with Windows 8.1, enumerating the most interesting parts:
\begin{enumerate}
\item The ``-FVE-FS-'' signature, which marks the beginning of an FVE block
\item The Windows version number
\item The type and value of a VMK metadata entry
\item According to this value, the VMK has been encrypted using the user password authentication method
\item The salt of the VMK
\item According to this value, the type of VMK encryption is AES-CCM
\item Nonce
\item Message Authentication Code
\item Finally, the VMK
\end{enumerate}

Our attack executes the BitLocker decryption procedure with several performance improvements:
% an improved version of this original BitLocker decryption procedure; in order to speed up the computation:
\begin{itemize}
\item The code has been optimized for NVIDIA GPUs (using the CUDA environment) (Section \ref{sec:bitcracker})
\item We introduced a preprocessing step, before starting the main attack, to store in memory useful information for the SHA-256 based main loop (Section \ref{sec:wwords})
\item We found a way to remove the final MAC computation and comparison (Section \ref{sec:metadata}).
\end{itemize}

\subsection{Recovery Password VMK Decryption Procedure\label{sec:rec_vmkdecryptionprocedure}}

As above mentioned, the recovery password is a kind of passe-partout for all the authentication methods.
According to \cite{bib:implbitlocker}, the recovery password is a 48-digit number composed by eight groups of six digits; each group of six digits must be divisible by eleven and must be less than 720896. Finally, the sixth digit in each group is a checksum digit.
For instance, a valid recovery password is:
\[\text{236808-089419-192665-495704-618299-073414-538373-542366}\]

The number of all possible recovery password candidates is huge, thus building the entire dictionary would require too much storage.

The algorithm used by BitLocker to encrypt a device using the recovery password
is similar to the user password one with few differences during the initial SHA-256 application.

\begin{figure}[h]
\begin{center}
\includegraphics[width=\linewidth]{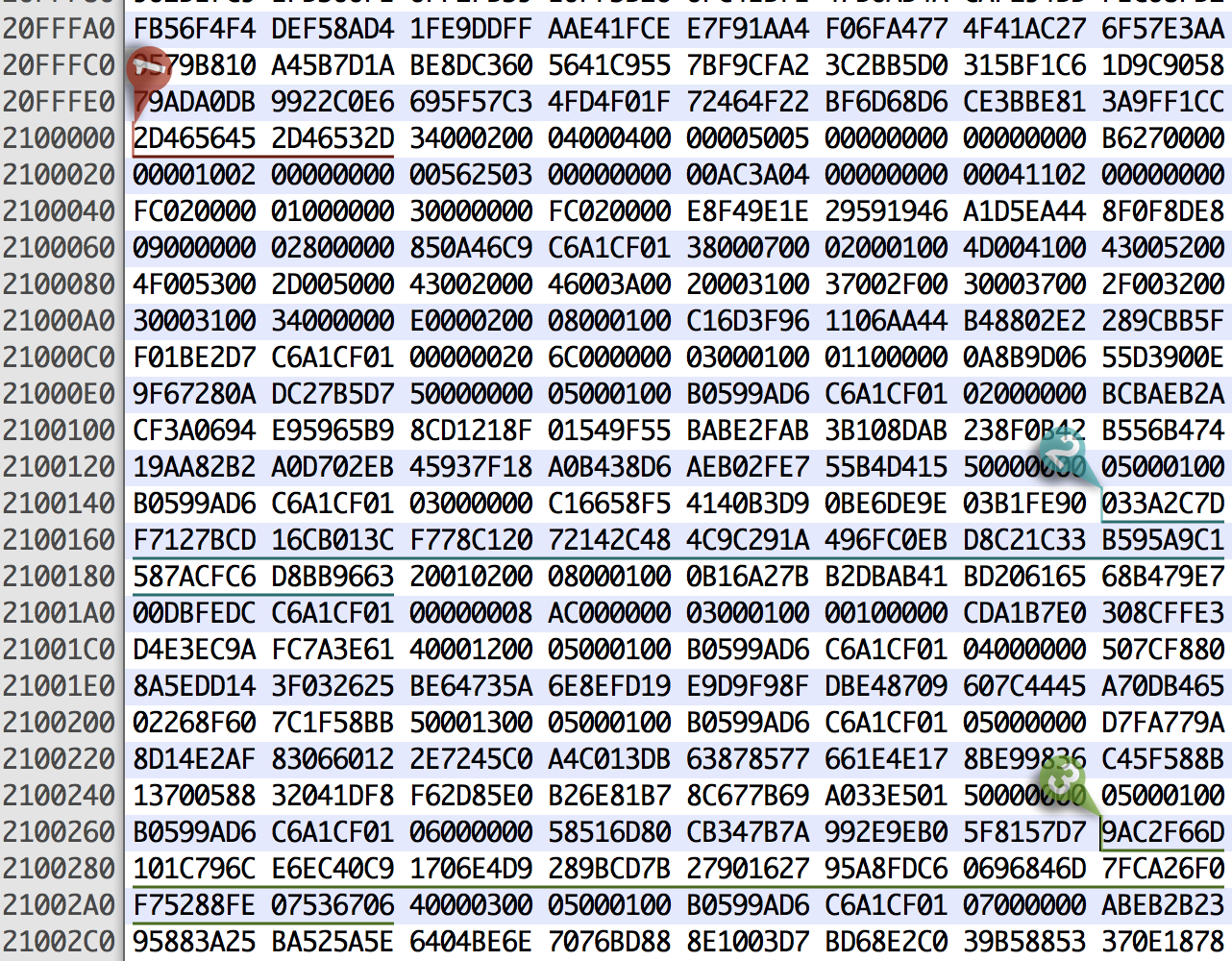}
\caption{FVE metadata block, BitLocker Windows 8.1, two Volume Master Keys\label{fig:fvevmk_recovery}}
\end{center}
\end{figure}

Figure \ref{fig:fvevmk_recovery} extends the Figure \ref{fig:fvevmk}; in the same FVE block (initial signature at mark number 1) there are 2 encrypted
VMKs: the first one (VMK\_U) is encrypted with the user password (mark number 2)
and the second one (VMK\_R) with the recovery password (mark number 3). The hardest part of the recovery password attack is to find the VMK\_R in the FVE blocks:
each authentication method (i.e. TPM, TPM+PIN, user password, smart card, etc..) has its
own FVE block format and each one stores the VMK\_R into a different index. Currently, we are able to find VMK\_R in case of devices encrypted with user password and smart card while TPM and TPM+PIN methods obfuscate the interesting part of the FVE block. BitCracker performance in case of recovery password attack is similar to the performance in case of user password; for this reason, during the rest of this paper, we report only about the performance of user password attacks.

\section{BitCracker \label{sec:bitcracker}}

Our software, named \textit{BitCracker}, aims at finding
(starting from a dictionary) the key of a memory unit encrypted
using the password authentication method of BitLocker.
To achieve that goal, BitCracker uses GPUs (\emph{Graphics Processing
  Units} \footnote{https://it.wikipedia.org/wiki/Graphics\_Processing\_Unit}) to execute the algorithm described in Section
\ref{sec:finalarchitecture} according to the Single Program, Multiple
Data (SPMD) paradigm: all threads execute same instructions on
different test passwords. In this Section we describe more in details the
behaviour of our algorithm and several optimizations we introduced to enhance its performance.

\subsection{SHA-256\label{sec:sha256}}

%SHA-2 (Secure Hash Algorithm) is a set of cryptographic hash functions
%designed by the NSA (U.S. National Security Agency). The SHA-2 family
%consists of six hash functions with digests (hash values) that are
%224, 256, 384 or 512 bits long.

As described in section \ref{sec:vmkdecryptionprocedure},
SHA-256 is widely used in the BitLocker decryption scheme to obtain the
intermediate key from the user password, so we focused our work in the
  improvement of its performance on GPU.
Algorithm \ref{algo:sha256standardalgo} presents a brief pseudo-algorithm of SHA-256 standard (for a full description, see \cite{bib:sha256standard}) that is necessary for
a better understanding of the work we did. It is apparent that:

\iffalse
\begin{enumerate}
\item Padding the input message. It must be a multiple of 512 bits (64
  bytes) padded in the following way:
\[
  M = \overbrace{
    \underbrace{m}_\text{original message} ||
    \underbrace{1}_\text{single bit 1} ||
    \underbrace{0}_\text{padding bits 0} ||
     \underbrace{size(m)}_\text{64 bits size}
   }^\text{padded message as multiple of 512 bits}
 \]
\item Parsing the padded message into N blocks of 64 bytes $M^{1}, M^{2}, .... ,M^{N}$
\item Setting the initial hash value $H^{0}_{0}, H^{0}_{1}, ... , H^{0}_{7} $ with constant values (see standard \cite{bib:sha256standard}).
\item Each message block $ M^{1}, M^{2}, .... , M^{N} $, is processed
  in order, using steps in Algorithm \ref{algo:sha256standardalgo}
\item After processing $ M^{N}$, the resulting 256-bit message digest of the
  original message M, is a chain of partial hash: $H^{N}_{0} ||
  H^{N}_{1} || H^{N}_{2} || H^{N}_{3} || H^{N}_{4} || H^{N}_{5} ||
  H^{N}_{6} ||  H^{N}_{7}$
\end{enumerate}
\fi

\begin{algorithm}
\small
\caption{SHA-256 standard algorithm}
\label{algo:sha256standardalgo}
\begin{algorithmic}[1]
\State Define Ch(x,y,z) = $(x \wedge y) \oplus (\lnot x \wedge z)$
\State Define Maj(x,y,z) = $(x \wedge y) \oplus (x \wedge z) \oplus (y \wedge z)$
\State Define $ROTR^{n}$(x) = $(x >> n) \vee (x << w-n)$ with $0 \leq n < w, w=32$
\State Define $SHR^{n}$(x) = $(x >> n)$
\State Define $ \sigma^{256}_{0}=ROTR^{7}(x) \oplus ROTR^{18}(x) \oplus SHR^{3}(x)$
\State Define $ \sigma^{256}_{1}=ROTR^{17}(x) \oplus ROTR^{19}(x) \oplus SHR^{10}(x)$
\State Define $M^{i}_{t}$ as the t-th 32-bit word belonging to the i-th 512-bit block of the padded message
\State Define K[64] as an array of constants
\State
\For{$i$ = 1 to $N$ }
\State Prepare the message schedule { $W_{t}$ }
\[ W_{t} =
  \begin{cases}
    M^{i}_{t}       & \quad \text{if } 0\le t \le 15\\
    \sigma^{256}_{1} (W_{t-2}) +  W_{t-7} + \sigma^{256}_{0} (W_{t-15}) +  W_{t-16}  & \quad \text{if } 16\le t \le 63\\
  \end{cases}
\]
\State Initialize the 8 working variables: a,b,c,d,e,f,g,h with the $(i-1)^{st}$ hash value:
\[ a = H^{i-1}_{0}; b = H^{i-1}_{1}; c=H^{i-1}_{2}; d=H^{i-1}_{3}; \]
\[ e = H^{i-1}_{4}; f=H^{i-1}_{5}; g=H^{i-1}_{6}; h=H^{i-1}_{7} \]
\For{$t$ = 0 to 63 }
\State $T_{1}$ = h + $ [ROTR^{6}(e) \oplus ROTR^{11}(e) \oplus ROTR^{25}(e)] + Ch(e,f,g) + K_{t}^{256} + W_{t}$
\State $T_{2}$ = $ [ROTR^{2}(a) \oplus ROTR^{13}(a) \oplus ROTR^{22}(a)] + Maj(a,b,c)$
\State h=g; g=f; f=e;
\State e=d + $T_{1}$
\State d=c; c=b; b=a;
\State a=$T_{1}$ + $T_{2}$
\EndFor
\State Compute the $i^{th}$ intermediate hash value $H^{i}$
\[ H^{i}_{0}  = a + H^{i-1}_{0}; H^{i}_{1}  = b + H^{i-1}_{1}; \]
\[ H^{i}_{2}  = c + H^{i-1}_{2}; H^{i}_{3}  = d + H^{i-1}_{3}; \]
\[ H^{i}_{4}  = e + H^{i-1}_{4}; H^{i}_{5}  = f + H^{i-1}_{5}; \]
\[ H^{i}_{6}  = g + H^{i-1}_{6}; H^{i}_{7}  = h + H^{i-1}_{7}; \]
\EndFor
\end{algorithmic}
\end{algorithm}

%Looking at Algorithm \ref{algo:sha256standardalgo},

\begin{itemize}
\item first 16 words of W depends on the original message;
\item there are two loops, the first to compute the $M^{N}$ message
  blocks and the second one to compute the 64 rounds of SHA-256;
\item circular shift is widely used and aritmetic operations are
  only bitwise, sum, and subtraction.
\end{itemize}

\iffalse
\newpage

\subsubsection{Parallel Computation\label{sec:sha256parallel}}

Although the SHA-256 algorithm is inherently serial, we started with a
minimal parallel version using just two threads to compute one hash as
described in listing \ref{list:sha256parallelalgo}.
\begin{lstlisting}[ caption=SHA-256 with two threads, label=list:sha256parallelalgo]
if(threadID%2 == 0)
{
        T1 = h + ROTR(e) + Ch(e,f,g) + K[i] + w[i];
        h=g; g=f; f=e; e=T1+tmpD;
}
if(threadID%2 == 1)
{
        T2 = ROTR(a) + MAJ(a, b, c);
        d=c; c=b; b=a; a=T2;
}
__syncthreads();

if(threadID%2 == 0) a += T1;
if(threadID%2 == 1) tmpD = d;

__syncthreads();
\end{lstlisting}
That approach works, but it was actually slower than the same
algorithm executed by a single thread, due to branch predication which
doesn't allow the parallel execution of the two {\em if} statements.
So, we discarded that parallel option
and we came back to a serial (meaning one-thread-one-password)
implementation.

\fi

\subsection{CUDA optimizations\label{sec:sha256operations}}

Our first implementation of the SHA-256 algorithm, was a plain C implementation quite similar to the algorithm described in Section \ref{sec:sha256}:
\begin{lstlisting}[ caption=SHA-256 C implementation, label=list:sha256c]
for(i=0; i<64; i++){
        T1 = h + S1(e) + CH(e,f,g) + K[i] + W[i];
        T2 = S0(a) + MAJ(a,b,c);
        h = g; g = f;
        f = e; e = d + T1;
        d = c; c = b;
        b = a; a = T1 + T2;
}
\end{lstlisting}

There are many assignments and loops with no GPU optimization.
We ran our first test on a NVIDIA GPU Tesla K80 for an initial
performance evaluation, reaching up to 80 passwords per second.

\iffalse
; a report of the timings is shown in Table \ref{tab:ctest}.

\begin{table}[H]
        \footnotesize
        \centering
    \begin{tabular}{|c|c|c|c|c|c|c|}
    \hline
        Blocks                       & Threads/Block               & Pwds/Thread          & Pwds/Kernel          & Seconds   & Pwds/Sec  \\ \hline
            1                             & 1.024                               & 8                                      & 8.192                   & 486              & 17                          \\
           6                             & 1.024                               &  8                                     & 49.152                 &  671                 &  73                         \\
          13                             & 1.024                               &  8                                     & 106.496               &  1055           & 100                              \\
    \hline
    \end{tabular}
    \caption{Initial performance without GPU improvements, NVIDIA GPU Tesla K80}
    \label{tab:ctest}
\end{table}

\fi

After that, inspired by Nayuki implementation \footnote{SHA-256 Nayuki implementation: \relax http://nayuki.eigenstate.org/page/fast-sha2-hashes-in-x86-assembly (Checked on May 20 2018)}, we wrote a code without loops (only inline
instructions), a lower number of operations, high throughput arithmetic instructions (see Section \ref{sec:cudaimpl}), removing all the useless aritmethics like index calculation ( $w-n$ ) or assignments (d=c; c=b; ...).\\
All GPU threads follow the same execution flow in a single SHA-256 execution: no shared memory is used and each thread
works on its subset of passwords stored in global memory; variables
used during execution are stored (as much as possible) in GPU
registers and all operations are performed inline, {\em i.e.,} there
are no loops (except for the loop of 0x100000 SHA-256) and very
few arrays index computations to limit {\em local} memory usage (more later).
The type of instructions used are: 32-bit integer add, 32-bit integer shift,
32-bit bitwise AND, OR, XOR (see Section \ref{sec:cudaimpl} for further details).

\begin{lstlisting}[caption=Final version of SHA-256, label=list:sha256final]
#define ROR6(x) (((x) << 26) | ((x) >> 6))
....
#define ROUND(a, b, c, d, e, f, g, h, i, k)
                h += (ROR6(e) ^ ROR11(e) ^ ROR25(e)) + (g ^ (e & (f ^ g))) + k + W[i];
                d += h;
                h += (ROR2(a) ^ ROR13(a) ^ ROR22(a)) + ((a & (b | c)) | (b & c));

function Sha256SingleExec() {
        calculateWwords(W, M);
....
        ROUND(a, b, c, d, e, f, g, h, 0, K[0])
        ROUND(h, a, b, c, d, e, f, g, 1, K[1])
        .....
        ROUND(b, c, d, e, f, g, h, a, 63, K[63])
...
        updateHashValue();
}
\end{lstlisting}

We reserve special attention to the usage of {GPU \em local} memory, because it can dramatically decrease performance.
Usually, if there is an array inside a kernel code
declared as a local variable but accessed dynamically inside the code
({\em i.e.,} array indexing is calculated at runtime) CUDA
stores that array in local memory (instead of using registers) and
this penalizes the performance since the {\em local} memory is
actually a part of the (slow) {\em global} memory of the GPU. We
reduced {\em local} memory usage as much as possible (we double-checked it
by looking at the low-level {\em PTX} code) by replacing index calculations
and loops.

During a second round of tests using the same NVIDIA GPU Tesla K80, we reached up to
103 passwords per second.

\iffalse
During a second round of tests, we evaluated our enhancements
running on the same GPU Tesla K80

In Table \ref{tab:wtest} we report the timings obtained with the
improved version of the SHA-256 algorithm and 0x100000 loop iterations.
In the 13 blocks test the gain is about 240 psw/sec.

\begin{table}[H]
        \footnotesize
        \centering
    \begin{tabular}{|c|c|c|c|c|c|c|}
    \hline
        Blocks                       & Threads/Block               & Pwds/Thread          & Pwds/Kernel          & Seconds   & Pwds/Sec  \\ \hline
            1                                    & 1.024                               & 8                                      & 8.192                   & 274              &  29                         \\
                6                                 & 1.024                               &  8                                     & 49.152                 & 275                  & 178                          \\
                13                               & 1.024                               &  8                                     & 106.496               & 306                   & 346                             \\
                13                               & 1.024                               &  128                       & 1.703.936            & 4.996            & 341                             \\
    \hline
    \end{tabular}
     \caption{Performance with GPU improvements, NVIDIA GPU Tesla K80}
    \label{tab:wtest}
\end{table}
\fi

After that, we focused on the VMK decryption algorithm (Figure \ref{fig:vmkdecrypt1}):
in Section \ref{sec:wwords} we describe our enhancement during the second SHA-256 in each iteration of the
main loop while in Section \ref{sec:metadata} we explain how the final comparison of the MAC can be easily removed.
%we removed the MAC comparison (section \ref{sec:metadata}) in order to save as much time as possible during the main attack.

\iffalse
with operations customized for different generations of GPU (``compute capabilities'' in CUDA jargon). In
particular, we developed two versions: the first one for Kepler
architecture (3.x compute capability) and the second one for Maxwell
architecture (5.x compute capability). For further details, see
Section \ref{sec:cudaimpl}.
\fi

\subsection{First improvement: W Words\label{sec:wwords}}

The most time-consuming part of the decryption algorithm is
the loop of 0x100000 (1.048.576) SHA-256 operations, since a single
hash involves many arithmetic operations. Moreover, during each
iteration, the SHA-256 algorithm is applied twice to the 128 byte structure \emph{bitlockerMessage}
(Section \ref{sec:vmkdecryptionprocedure}) which is composed by several fields
as shown in Table \ref{tab:bitlockermessage}.

\begin{table}[h!]
  \begin{center}
\scriptsize
\begin{tabular}{|*{6}{c|}} 
\hline
\multicolumn{2}{|c}{64-byte block \#1} & \multicolumn{4}{|c|}{64-byte block \#2} \\ \hline
32 bytes & 32 bytes & 16 bytes & 8 bytes & 32 bytes & 8 bytes\\ \hline
updated\_hash & password\_hash & salt & hash\_count & padding & message size\\ \hline
variable & fixed & \shortstack{fixed for each \\ encrypted unit} &  \shortstack{between 0 and \\ 0x100000} & fixed & fixed to 88\\ \hline
\end{tabular}
\caption{BitLocker SHA-256 message\label{tab:bitlockermessage}}
  \end{center}
\end{table}
\normalsize

As mentioned above, the first 16 $W$ words depend on the original message and
the others on the first 16 words. Therefore, looking at the message in
Figure \ref{tab:bitlockermessage} we were able to compute all possible
$W$ words useful for the SHA-256 of the second block of the message at
each iteration in the loop, with no need to repeat many arithmetic
operations during each iteration. Indeed, since for each encrypted
memory unit, salt, padding and message size are always the same
and {\tt hash$\_$count} is a number between 0 and (0x100000-1), we can precompute all the W words, that are:
\[ 1.048.576 * 64 =  67.108.864 \ words * 4 \ byte \simeq 256 Mb \]

This kind of improvement is specific for BitLocker (precomputation
  can be done if there is a part of the input message that is known ahead of time) and cannot be applied to a general SHA-256 implementation.

  In the beginning, we stored pre-computed $W$ words in global memory,
  but we found that texture memory (due to the texture caching
  capabilities) could improve timings (see Section
  \ref{sec:cudaimpl}). For the first block, we can not precompute
  anything, because \emph{updateHash} changes at every iteration.

Thanks to this improvement we reduced the CUDA registers pressure and usage,
being able to use 64 registers for each CUDA thread and then 1024 threads for
each CUDA block (about 100\% occupancy).
We measure a performance enhancement, reaching up to 340 passwords per second.

  %To save some registers, we used an array of 32 bytes instead of 64 bytes
  %to store $W$ words during this first hash procedure.

\subsection{Second improvement: MAC comparison\label{sec:metadata}}

\iffalse

The structure of a single metadata is described in Table \ref{tab:metadatastructure}:

\begin{table}[H]
        \small
        \centering
    \begin{tabular}{|c|c|c|c|c|c|c|}
    \hline
    Offeset             & Bytes      & Description \\ \hline
    0                  & 2            & Byte size of metadata \\
    2                  & 2            & Type of metadata \\
    4                  & 2            & Value of metadata \\
    6                  & 2            & Version \\
    8                  &               & Data (Body of metadata) \\
    \hline
    \end{tabular}
        \caption{Metadata Structure}
    \label{tab:metadatastructure}
\end{table}

In case of a VMK encrypted with a user password, the
type of metadata is equal to \emph{0x0002} and its value is equal to \emph{0x0008}.
After the first 12 bytes inside the body of the VMK metadata there are: % (Figure \ref{fig:vmkmetadata}):
\begin{itemize}
\item \textbf{Salt}: added to the user password (16 bytes)
\item \textbf{Size}: VMK size (2 bytes)
\item \textbf{Type}: encryption type (2 bytes)
\item \textbf{Nonce}
\item \textbf{MAC}: Message Authentication Code
\item \textbf{VMK} encrypted
\end{itemize}

\begin{figure}[h]
\begin{center}
\includegraphics[width=\linewidth]{vmkMetadata.png}
\caption{VMK metadata. Blue digits are sizes, purple are types of
  encryption, yellow are the salt, red are the nonce, green are the
  MAC and grey are the keys\label{fig:vmkmetadata}.}
\end{center}
\end{figure}
\fi

During our analysis of the decrypted VMK's structure, using different
Windows versions (7, 8.1 and 10) and a number of encrypted devices, we noticed
several interesting facts:
\begin{enumerate}
\item The size of the VMK is always 44 bytes
\item First 12 bytes of decrypted VMK (Table \ref{tab:vmk_header}) hold information about the key
        \begin{itemize}
        \item First 2 bytes are the size of VMK, that is always 44 (0x002c)
            \item Bytes 4 and 5 are the \textit{version} number, always equal to 1
                \item Byte 8 and 9 are the type of encryption.
          In case of user password, BitLocker
          always uses AES-CCM with a 256 bit key. So, according to the Microsoft
          standard, this value is between 0x2000 and 0x2005
    \end{itemize}
\item Remaining 32 bytes are the key
\end{enumerate}

\begin{table}[H]
        \centering
    \begin{tabular}{|c|c|c|c|c|c|c|c|c|c|c|c|c|}
        Byte                       & 0 & 1 & 2 & 3 & 4 & 5 & 6 & 7 & 8 & 9 & 10 & 11 \\ \hline
        Value                       & 2c & 00 & 00 & 00 & 01 & 00 & 00 & 00 & 03 & 20 & 00 & 00 \\
    \end{tabular}
    \caption{Example of initial 12 bytes of VMK decryption key}
    \label{tab:vmk_header}
\end{table}

\begin{figure}[h]
\begin{center}
\includegraphics[width=\linewidth]{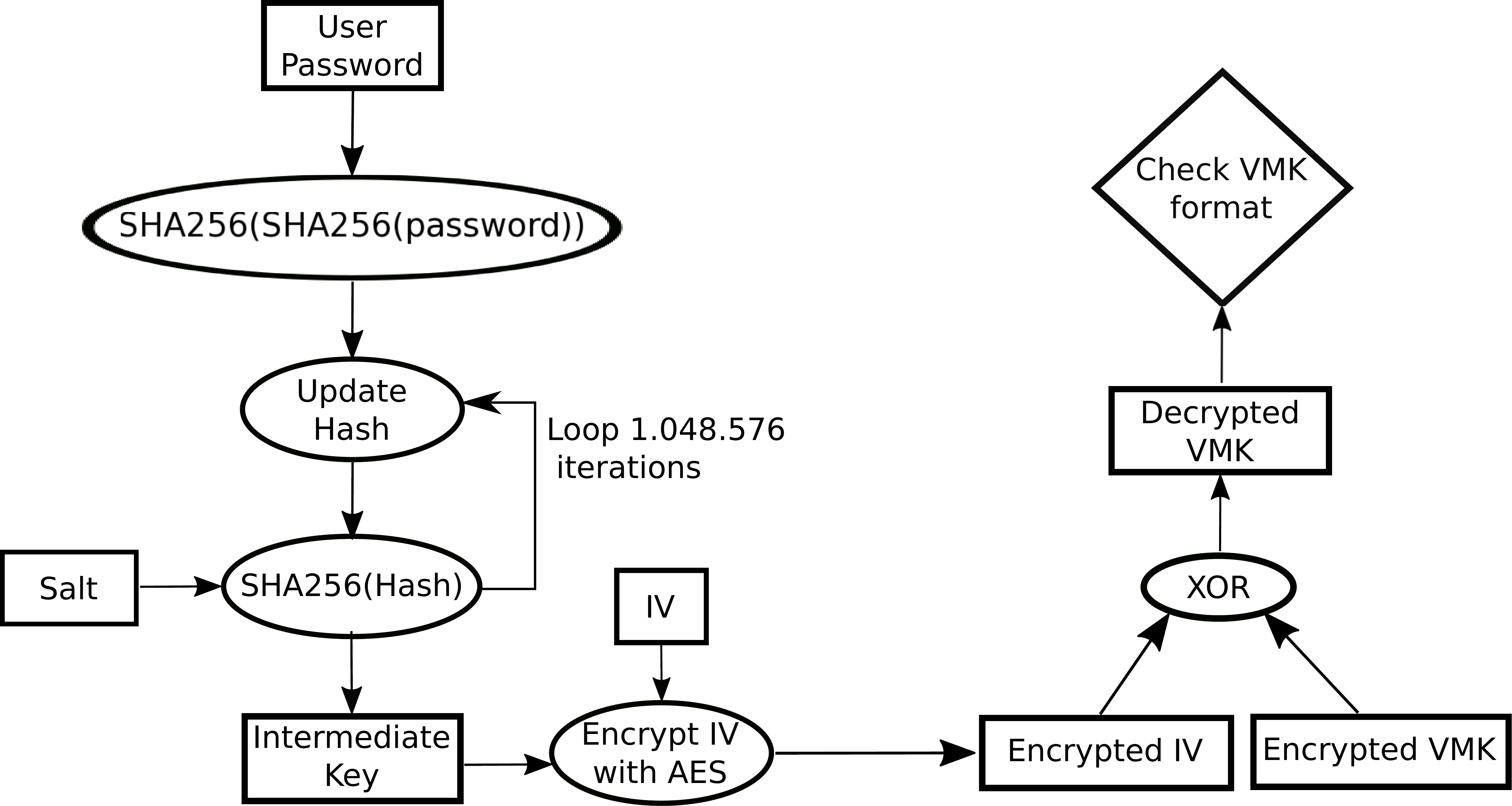}
\caption{VMK decryption procedure improved\label{fig:vmkdecrypt2}}
\end{center}
\end{figure}

Following those considerations, we removed the MAC test doing a simple check on the initial
 12 bytes of the decrypted VMK, as shown in Figure \ref{fig:vmkdecrypt2}.

On our GPU Tesla K80 performance reached 385 passwords per second (see Section \ref{sec:cudaimpl} for further details).

To check the reliability of our solution, we tested BitCracker with
  several storage devices (both internal and USB-connected
  hard disks) encrypted by using passwords having between 8 and 16
  characters under Windows 7 Enterprise Edition, Windows 7
  Ultimate Edition and Windows 8 Pro N and Windows 10 Enterprise Edition
  (testing both BitLocker's compatible and non compatible modes)
  \footnote{Recently Microsoft introduced the BitLocker "Not Compatible" encryption mode in Windows 10: sectors of the memory device are encrypted with XTS-AES instead of AES-CCM. This change
  doesn't affect BitCracker algorithm because there isn't any difference in the decryption procedure of the VMK.}.

%We carried out an extensive set of tests using different Windows versions to encrypt memory devices performing this alternative and faster VMK decryption procedure, checking
% different combinations of the properties listed above.

Although BitCracker always returned the correct output, some false positive may occur with this improved VMK check;
 for this reason BitCracker can be executed in 2 different modes: with (slower solution) or without (faster solution) the MAC comparison .

 \iffalse
 We got some false positive without the \textit{version} check,
 due to the type of encryption range values (between 0x2000 and 0x02005).
 Although it isn't specified in the standard, we observed that in case of the correct password the type of encryption AES-CCM
 is always 0x2003; we can't guarantee that this check doesn't lead to false negatives.
 Thus the final VMK check is:

 \begin{itemize}
 \item Required:
         \begin{itemize}
                 \item Bytes 0 and 1 must be equal to 0x002c
                 \item Bytes 4 and 5 must be equal to 0x0001
                 \item Bytes 8 and 9 must be $\leq$ than 0x2005
        \end{itemize}
 \item Optional:
  \begin{itemize}
                 \item Bytes 8 and 9 must be equal to 0x2003
 \end{itemize}
 \end{itemize}
  \fi

\subsection{Final Architecture\label{sec:finalarchitecture}}

We implemented the final solution, shown in Algorithm \ref{algo:bitcrackersha256}, using CUDA and OpenCL doing an extensive performance analysis in Sections \ref{sec:cudaimpl} and \ref{sec:opencl}.

%The final solution is shown in Algorithm \ref{algo:bitcrackersha256} and its performance is reported in Section \ref{sec:cudaimpl}.
\begin{algorithm}
\caption{BitCracker Kernel}
\small
\label{algo:bitcrackersha256}
\begin{algorithmic}[1]
\Function{BitCrackerKernel}{password, WTextureWords, IV, VMK}
        \State hash = Sha256SingleExec(Sha256SingleExec(password));
        \For{$i$ = 1 to 0x100000 }
                \State SetStartValue(a,b,c,d,e,f,g,h);
                \State \Comment /* SHA-256 on the first message block  */
                \State Compute first 32 W words, depending on $hash_{i-i}$
                \State Exec first 32 rounds
                \State Compute second 32 W words
                \State Exec second 32 rounds
                \State Update $hash_{i}$ value
                \State \Comment /* SHA-256 on the second message block  */
                \State Exec 64 rounds, reading W blocks from WTextureWords
                \State Update $hash_{i}$ value
        \EndFor
        \State Crypt IV with final hash value as key of AES
        \State VMKDecrypted = XOR(IVCrypted, VMK)
        \If { checkVMK(VMKDecrypted) }
                \State return true;
        \EndIf
\EndFunction
\end{algorithmic}
\end{algorithm}

\begin{figure}[h]
\begin{center}
\includegraphics[width=\linewidth]{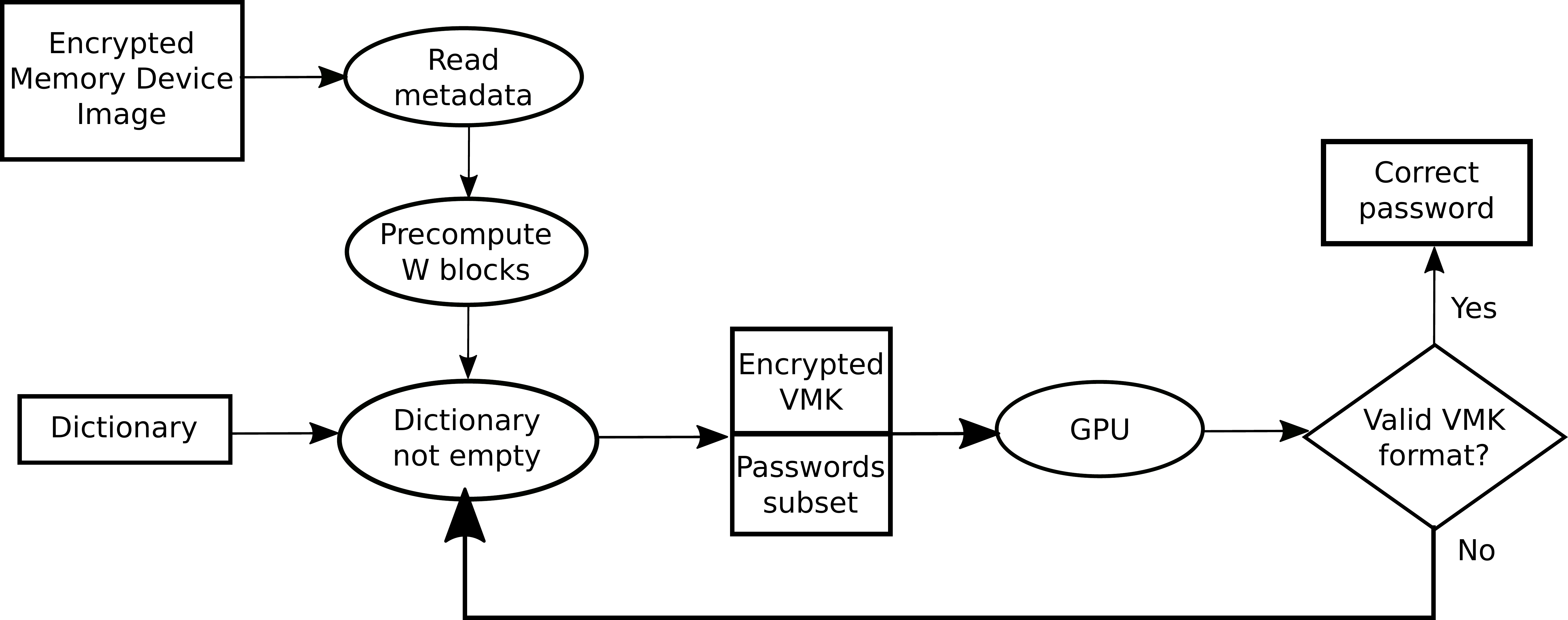}
\caption{BitCracker general algorithm\label{fig:bitlockermonogpu}}
\end{center}
\end{figure}

As described in Section \ref{sec:vmkdecryptionprocedure},
the main SHA-256 loop produces the intermediate key, that must be used
as an AES key to encrypt the IV in order to decrypt the VMK.
We have implemented our own AES version customized for GPU environment \cite{bib:Dcrack},
used in line 15 of Algorithm \ref{algo:bitcrackersha256}.
In Figure \ref{fig:bitlockermonogpu} we represent the entire BitCracker's general procedure outside of the GPU kernels.

\section{CUDA implementation performance\label{sec:cudaimpl}}

In this Section we present the results of benchmarking activities of our stand-alone CUDA implementation of BitCracker with the improvements described in previous sections.
We used several NVIDIA GPUs whose features are summarized in Table \ref{tab:nvidiagpus} \footnote{\textit{CC} is Compute Capability while \textit{SM} is Stream Multiprocessors}.
  \begin{table}[H]
        \small
        \centering
    \begin{tabular}{|c|c|c|c|c|c|c|c|}
    \hline
    Acronim             & Name                                  & Arch             & CC                    & \# SM  & SM Clock   & CUDA          \\ \hline
    GFT                 & GeForce Titan                 & Kepler                & 3.5           & 14     & 836 MHz       & 7.0           \\ \hline
    GTK80            & Tesla K80                        & Kepler                & 3.5           & 13     & 875 MHz       & 7.0/7.5           \\ \hline
    GFTX              & GeForce Titan X                 & Maxwell               & 5.2           & 24     & 1001 MHz      &  7.5          \\ \hline
    GTP100          & Tesla P100                        & Pascal                & 6.1           & 56     & 1189 MHz      &  8.0          \\ \hline
    GTV100          & Tesla V100                        & Volta                  & 7.0           & 80     & 1290 MHz     &  9.0          \\ \hline
    \end{tabular}
    \caption{NVIDIA GPUs used for bench}
    \label{tab:nvidiagpus}
\end{table}

During the following tests we always set the number of CUDA blocks to the maximum number of SM allowed by the GPU architecture: further increasing this number does not improve performance. The number of CUDA threads per block is always 1024 because each thread requires no more than 64 registers (we reached the maximum occupancy).

\subsection{Kepler Architecture}

We started to benchmark our final improved solution on the {\em Kepler} architecture using GPUs GTK80 (Table \ref{tab:teslatest}) and GFT (Table \ref{tab:geforcetest}).

  \begin{table}[H]
        \footnotesize
        \centering
    \begin{tabular}{|c|c|c|c|c|c|}
    \hline
    Blocks                       & Threads/Block               & Pwds/Thread          & Pwds/Kernel          & Seconds   & Pwds/Sec  \\ \hline
      1                          & 1.024                                        &  1                                                    & 1.024                           & 30                          &  33      \\
      1                          & 1.024                                        &  8                                            & 8.192                                 & 245                           &  33      \\
      2                          & 1.024                                        &  8                                            & 16.384                        & 247                   &  66      \\
      4                          & 1.024                                        &  8                                            & 32.768                        & 248             &  132     \\
      8                          & 1.024                                        &  8                                            & 65.536                        & 253            &  258     \\
      13                         & 1.024                                         &  8                                           & 106.496                       & 276            & 385      \\
%      \hline
%      13                         & 1.024                         &  128                  & 1.703.936             & 4.594         & 370     \\
    \hline
    \end{tabular}
    \caption{GTK80 benchmarks}
    \label{tab:teslatest}
\end{table}

\begin{table}[H]
        \footnotesize
        \centering
    \begin{tabular}{|c|c|c|c|c|c|}
    \hline
        Blocks         & Threads/Block               & Pwds/Thread          & Pwds/Kernel          & Seconds   & Pwds/Sec  \\ \hline
        1              & 1.024                         &  1                    & 1.024                 & 39            &  25       \\
        1              & 1.024                         &  8                    & 8.192                 & 318           &  25       \\
        14             & 1.024                         &  8                    & 114.688                     & 364           & 314         \\
%        \hline
%        14            & 1.024                          &  128                     & 1.835.008                    & 6051         &  303        \\
    \hline
    \end{tabular}
    \caption{GFT benchmarks}
    \label{tab:geforcetest}
\end{table}

The more the input grows, the better BitCracker performs. Increasing
the number of blocks, each one with the same number of passwords per
thread ({\em i.e.,} 8), leads to a better performance since the kernel
launching overhead (that is basically constant) is distributed among
more blocks.

%about $20\%$

The GTK80 appears to be faster with respect to the GTF. For instance, in the 8 passwords
per thread last test-case the GTK80 is able to check $\sim 70$
more passwords per second than the GFT even if it has one
less multiprocessor. To shed some light on this
difference, we report in Table \ref{tab:instructioniteration} a
summary of instructions types and occurrences in one out of 0x100000
iterations of the SHA-256 loop.

\begin{table}[H]
        \centering
        \footnotesize
    \begin{tabular}{|c|c|c|c|c|c|c|}
    \hline
    Instruction                 & Throughput GTK80              &       Throughput GTF           & \# Occurrence      \\ \hline
    32-bit integer add          &       160                             &       160                         & 1.121                    \\
    32-bit integer shift        &       64                      &       32                              &  2.016                    \\
    32-bit bitwise                      &       160                      &       160                         & 1.600                      \\
    \hline
    \end{tabular}
    \caption{Instructions throughput}
    \label{tab:instructioniteration}
\end{table}

As described in the NVIDIA developer zone \footnote{NVIDIA Developer Zone: \relax https://technet.microsoft.com/en-us/library/cc162804.aspx}:
\begin{itemize}
\item to maximize instruction throughput, the application should minimize the use of arithmetic instructions with low throughput;
\item throughput is given in number of operations per clock cycle per multiprocessor. For a warp size of 32, one instruction corresponds to 32 operations, so if N is the number of operations per clock cycle, the instruction throughput is N/32 instructions per clock cycle.
\end{itemize}
Since the 32-bit integer shift is the most used instruction (due to the
circular shift, see Section \ref{sec:bitcracker}) the fact that
its throughput in the GTK80 is twice than in the GFT could explain the significant difference in performance.
Due to the relevance of the shift instruction, the circular shift is translated inside
the PTX code in a \emph{funnel shift} operation (faster than
\emph{regular shift}) even if this is not explicitly specified in the CUDA code. That is, an instruction like:
\[ ((x) \ << 26) \ | \ ((x) \ >> \ 6)) \]
becomes, in PTX code:
\[ shf.l.wrap.b32 \ \%r8165, \ \%r8152, \ \%r8152, \ 26 \]

We limit the number of registers for a single thread to 64. This
choice makes it possible to have 1.024 threads for each block since
each Stream Multiprocessor for Kepler and Maxwell GPU has 65536 32-bit
registers. On the GTK80 there is no memory spilling while on the GFT we have
just 24 bytes of memory spilling (handled
by the L1 cache). Nevertheless, this configuration offers the best
performance. The occupancy level is close to 100\% provided that there
are enough passwords in input (shared memory is not a limiting factor
since we do not use it).

\iffalse
On the GTK80 and GFTX there is no memory spilling.
\fi

%\subsubsection{Profiler Analysis\label{sec:profileranalysis}}

In Table \ref{tab:nvprof} we report some of the metrics provided by the CUDA
profiler \verb|nvprof| running a synthetic test in which the number of loop iterations is limited to 65.536 to prevent
the overflow of some counters that occurred in the full 0x100000 iterations execution:

\begin{table}[H]
        \small
        \centering
    \begin{tabular}{|l|l|}
    \hline
    Instructions issued                     & 8.072.134.346 \\ \hline
    Instruction replay overhead                 & 0,000003      \\ \hline
    Global memory load                      & 2.208         \\ \hline
    Global memory store trans.                  & 1                     \\ \hline
    Local memory load trans.                            & 0             \\ \hline
    Local memory store trans.               & 0                 \\ \hline
    Arithmetic Function Unit Utilization    & High(9)       \\ \hline
    Texture cache hit rate                  & 99.80\%           \\ \hline
    \end{tabular}
    \caption{nvprof Metrics on GTK80}
    \label{tab:nvprof}
\end{table}

The number of {\em instructions issued} is greater than the number of {\em global memory transactions}, so BitCracker can be considered, somehow,
instructions limited (this is not surprising since there are very few global memory load/store operations).
The metrics about the texture cache confirms that the use of texture
memory for $W$ words (instead of global memory) is optimal.
There is no instruction serialization inside a single warp (usually
serialization is due to memory conflicts) because {\em instruction replay overhead} metrics is very close to 0.
Finally, there are no local memory transactions due to our optimization work described in Section \ref{sec:sha256operations}.\\

\subsection{Maxwell Architecture}\label{sec:maxwell}

%In the Maxwell architecture, 3 additions or 3 bitwise
%instructions can be merged in a single instruction according to new
%PTX opcodes introduced since CUDA 7.5.
In Table \ref{tab:geforcetestx} we present the same benchmarks of the previous Section executed on the GFTX, using CC3.5 and CC5.2 (both available on the GPU).

\begin{table}[H]
        \scriptsize
        \centering
    \begin{tabular}{|c|c|c|c|c|c|c|}
    \hline
    CC          &Blocks         & Threads/Block               & Pwds/Thread          & Pwds/Kernel          & Seconds   & Pwds/Sec  \\ \hline
    3.5         &1              & 1.024                         &  1                    & 1.024                 & 24            &  42       \\
    3.5         &1              & 1.024                         &  8                    & 8.192                 & 191           &  42       \\
    3.5         &24             & 1.024                         &  8                    & 196.608               &  212          &  925        \\
        \hline
    3.5         &24             & 1.024                         &  128                  & 3.145.728             &  3496         &  900       \\
    \hline
    5.2         &1              & 1.024                         &  1                    & 1.024                 & 23            &  44       \\
    5.2         &1              & 1.024                         &  8                    & 8.192                 & 188           &  43       \\
    5.2         &24             & 1.024                         &  8                    & 196.608               & 210           &  933        \\
        \hline
    5.2         &24            & 1.024                          &  128                   & 3.145.728            & 3369          &  933       \\
    \hline
    \end{tabular}
    \caption{GFTX benchmarks, CC3.5 and CC5.2}
    \label{tab:geforcetestx}
\end{table}

\iffalse
To make a comparison between those two architectures we ran some tests using the 3.5 compute capability code on
the Maxwell card ({\em i.e.,} GPU Titan X) so disabling optimizations for the 5.x compute capability.
Looking at the first part of Table \ref{tab:geforcetestx},
\fi
It is worth to note that performance improves both due to the higher number of
multiprocessors available in the new generation of NVIDIA cards and
for the enhancements in integer instructions throughput \footnote{NVIDIA Developer Zone Maxwell: \relax https://developer.nvidia.com/maxwell-compute-architecture}. This confirms that a well-tuned CUDA code
can benefit from new features with a very limited effort.

With CUDA 7.5 NVIDIA released new PTX instructions like
  \emph{IADD3} and \emph{LOP3}, that support a range of 3-operand logic operations,
  such as $ (A \& B \& C)$, $(A \& B \& \neg C)$, $(A \& B \lor C)$ and so on.
  According to the release notes in \footnote{NVIDIA Devblog: https://devblogs.nvidia.com/parallelforall/new-features-cuda-7-5}, those instructions are fully supported on GPUs with CC 5.0 or greater, while on a Kepler architecture they are simulated; therefore we optimized BitCracker code for Maxwell cards.

%Therefore we optimized BitCracker code for Maxwell cards
%with CC 5.2 by using new PTX instructions like
%  \emph{IADD3} and \emph{LOP3}, that support a range of 3-operand logic operations,
%  such as $ (A \& B \& C), (A \& B \& \neg C), (A \& B \lor C)$ and so on (see \cite{bib:devblog}).
In other words, the \emph{LOP3} instruction combines three operands according
to a Truth Table expressed as a hexadecimal number; in Table \ref{tab:xortruthtable}
there is an example of a XOR Truth Table. See \footnote{NVIDIA PTX LOP3: http://docs.nvidia.com/cuda/parallel-thread-execution/\# logic-and-shift-instructions-lop3} for further details and examples.

\begin{table}[H]
        \footnotesize
        \centering
    \begin{tabular}{|c|c|c|c|}
    \hline
    a   &b      &c              &result                 \\ \hline
    0   &0      &0              &0                      \\
    0   &0      &1              &1                      \\
    0   &1      &0              &1                      \\
    0   &1      &1              &0                      \\
    1   &0      &0              &1                      \\
    1   &0      &1              &0                      \\
    1   &1      &0              &0                      \\
    1   &1      &1              &1                      \\
    \hline
                &       &               &$0x96$                 \\
        \hline
    \end{tabular}
    \caption{XOR Truth Table}
    \label{tab:xortruthtable}
\end{table}

Looking at Table \ref{tab:geforcetestx}, Maxwell card performance increases
when using 5.2 compute capability with an intensive use of the above mentioned PTX
instructions for all the bitwise boolean operations involved in the SHA-256
algorithm described in \ref{sec:sha256operations}.

\iffalse
Nevertheless, as described in Section \ref{sec:vmkdecryptionprocedure}, the
bottleneck remains the 0x100000 SHA-256 iterations. To prove that, we
launched a modified BitCracker kernel without the main loop ({\em i.e.,} the
execution uses the output of the second SHA-256 to compute AES) using 13 CUDA blocks with 1.024 threads, each one checking 128 password: the execution time was about 0.127 sec, proving that the first two SHA-256 and the AES computations require a very limited amount of time.
\fi

\subsection{Pascal architecture}\label{sec:pascal}

In Table \ref{tab:tesla_pascal}, we summarize our benchmarks on GTP100.
The performance improvement is close to a  $\times~2$ factor with respect to
the Maxwell architecture even if the main advantage of the new
architecture ({\em i.e.}, the memory bandwidth that is about three times
higher with respect to the {\em Kepler} architecture) has limited
impact on a compute-intensive application like BitCracker.

\begin{table}[H]
        \scriptsize
        \centering
    \begin{tabular}{|c|c|c|c|c|c|c|}
    \hline
    CC          &Blocks    & Threads/Block               & Pwds/Thread          & Pwds/Kernel       & Seconds   & Pwds/Sec  \\ \hline
    6.1         &1         & 1.024                         &  1                    & 1.024              & 38            &   26       \\
    6.1         &56        & 1.024                         &  1                    & 57.344             & 40            & 1.418       \\
    6.1         &56        & 1.024                         &  8                    & 458.752            & 336           & 1.363         \\
    6.1         &56        & 1.024                         &  128                  & 7.340.032                  & 5444          & 1.348        \\
    \hline
    \end{tabular}
    \caption{GTP100 benchmark}
    \label{tab:tesla_pascal}
\end{table}

\subsection{Volta architecture}\label{sec:volta}

In Table \ref{tab:tesla_volta}, we summarize our benchmarks on GTV100.
The performance improvement is more than $\times~2$ factor with respect to
the Pascal architecture.

\begin{table}[H]
        \scriptsize
        \centering
    \begin{tabular}{|c|c|c|c|c|c|c|}
    \hline
    CC          &Blocks    & Threads/Block     & Pwds/Thread    & Pwds/Kernel       & Seconds   & Pwds/Sec  \\ \hline
    7.0         & 1             & 1.024                         &  1                    & 1.024                         & 24            &   41                          \\
    7.0         & 80            & 1.024                         &  1                    & 81.920                        & 25            & 3.252       \\
    7.0         & 80            & 1.024                         &  8                    & 655.360               & 210          & 3.107         \\
    \hline
    \end{tabular}
    \caption{GTV100 benchmark}
    \label{tab:tesla_volta}
\end{table}

\section{OpenCL Implementation}\label{sec:opencl}

In order to make BitCracker available also to non-NVIDIA GPUs, we
developed an OpenCL implementation.

In Table \ref{tab:opencl_benchmarks_amd} we report performance of BitCracker-OpenCL standalone version tested with an AMD Radeon HD 7990
Malta\footnote{by courtesy of the Openwall HPC Village\cite{bib:openwall}.}, using OpenCL version 1.2 (1800.5), confirming that BitCracker OpenCL is able to run on non-NVIDIA GPUs.

\begin{table}[H]
        \scriptsize
        \centering
    \begin{tabular}{|c|c|c|c|c|c|c|}
    \hline
    Work Groups    & Local Work Group size   & Pwds/Thread   & Pwds/Kernel & Seconds  & Pwds/Sec  \\ \hline
    32             & 256                     & 1                  & 8.192        &  41          & 197        \\
    64             & 256                     & 1                  & 16.384       &  68          & 241        \\
    32             & 256                     & 64               & 524.288      &  2657        & 197        \\
    64             & 256                     & 32               & 524.288      &  2175        & 241        \\
    64             & 256                                                & 48                    & 786.432              &  3263         & 241   \\
   \hline
    \end{tabular}
    \caption{Benchmarks OpenCL version, AMD Radeon HD 7990 Malta}
    \label{tab:opencl_benchmarks_amd}
\end{table}

\iffalse
For performance comparison with the CUDA implementation,
in Table \ref{tab:volta_opencl} we show that the OpenCL implementation is just
$5\%$ slower with respect to the CUDA original implementation (see
Table \ref{tab:tesla_volta}) .

\begin{table}[H]
        \scriptsize
        \centering
    \begin{tabular}{|c|c|c|c|c|c|c|}
    \hline
    Work Groups    & Threads/Block           & Pwds/Thread          & Pwds/Kernel          & Seconds   & Pwds/Sec  \\ \hline
    1                                   & 1.024                         &  1                                            & 12.288                        & 286            &  43                  \\
    80                                  & 1.024                         &  8                                            & 983.040               & 229                    & 3286       \\
    \hline
    \end{tabular}
    \caption{GTV100, OpenCL 1.2}
    \label{tab:volta_opencl}
\end{table}
\fi

%\iffalse
The results, reported in Table \ref{tab:geforcetestx_opencl}, show that the OpenCL implementation is just
$5\%$ slower with respect to the CUDA original implementation on the same platform (see Table \ref{tab:geforcetestx}).

\begin{table}[H]
        \scriptsize
        \centering
    \begin{tabular}{|c|c|c|c|c|c|c|}
    \hline
    Work Groups    & Threads/Block               & Pwds/Thread          & Pwds/Kernel          & Seconds   & Pwds/Sec  \\ \hline
    1              & 1.024                         &  1                    & 1.024                 & 23            &  44       \\
        24         & 1.024                         &  8                    & 196.608               & 220                    & 893         \\
    24             & 1.024                         &  128                  & 3.145.728             & 3555          &  884       \\
    \hline
    \end{tabular}
    \caption{GFTX, OpenCL 1.2}
    \label{tab:geforcetestx_opencl}
\end{table}
%\fi

\section{Performance comparison\label{sec:performance}}

It is possible to evaluate BitCracker's performance by looking at the
number of hashes per second that it computes (we recall that
the check of each password requires 2.097.154 hashes, as described in
Section \ref{sec:vmkdecryptionprocedure}).
The number of hashes {\em per} second that BitCracker is able to perform is summarized in Table \ref{tab:hashescomparison}\footnote{MH stands for MegaHashs} .

\begin{table}[H]
        \centering
    \begin{tabular}{|c|c|c|c|}
    \hline
    GPU                & Password/Sec   & Hash/Sec          \\ \hline
    GFT                & 303                    & 635 MH/s          \\
    GTK80              & 385                    & 807 MH/s          \\
    GFTX               & 933                    & 1.957 MH/s         \\
    GTP100             & 1.418                & 2.973 MH/s         \\
    GTV100             & 3.252               & 6.820 MH/s         \\
    \hline
    \end{tabular}
    \caption{BitCracker's hashes per second, CUDA implementation}
    \label{tab:hashescomparison}
\end{table}

\subsection{Hashcat comparison}\label{sec:hashcat}

To assess BitCracker performance, we carried out a comparison with
the SHA-256 format (-m 1400) Hashcat \cite{bib:hashcat} v4.1.0.
We highlight that this is not a completely fair comparison since Hashcat does
not execute exactly the same BitCracker's algorithm (BitCracker performs other operations beyond SHA-256) and it currently supports OpenCL only.
The test aims at providing an idea about the number of SHA256 that each one of them is able to
compute {\em per} second. We ran a test on the GTV100 using the following parameters:
\begin{itemize}
\item -m 1400 : Raw SHA-256 hash format
\item -a 3 : Mask attack
\item ?a?a?a?a?a?a?a?a?a : Mask to specify passwords of 8 characters
\item -d : specify the GTV100 device
\item -O and -w 3 as suggested by Hashcat itself to improve performance
\end{itemize}
The resulting number of hashes per second is 7590 MH/s that is comparable to BitCracker's best performance on the same GPU.\\

\subsection{John The Ripper}\label{sec:jtr_format}

In order to take advantage of their system of {\em rules} for wordlist
generation, our OpenCL implementation has been released as a John the Ripper (Jumbo version) \cite{bib:jtr} plugin (format name \textit{bitlocker-opencl});
the source code can be found here \cite{bib:jtr_github} whereas the wiki reference page is here \cite{bib:jtr_wiki}.
When running \textit{bitlocker-opencl} format, the John The Ripper internal engine auto-tunes all the OpenCL parameters (like local and global work groups). Running a simple test like:
\[ \text{./john --format=bitlocker-opencl --wordlist=wordlist.txt hashFile.txt} \]
we reached up to 3150 passwords/second on the GTV100 .

The John The Ripper team developed a CPU format of our attack which can be invoked using flag ``\text{\textit{--format=bitlocker}}''.
We reached up to 78 passwords/second on a CPU Intel(R) Xeon(R) v4, 2.20GHz.

\subsection{Performance overview}\label{sec:performance_overview}

\begin{figure}[H]
\begin{center}
\includegraphics[width=\linewidth]{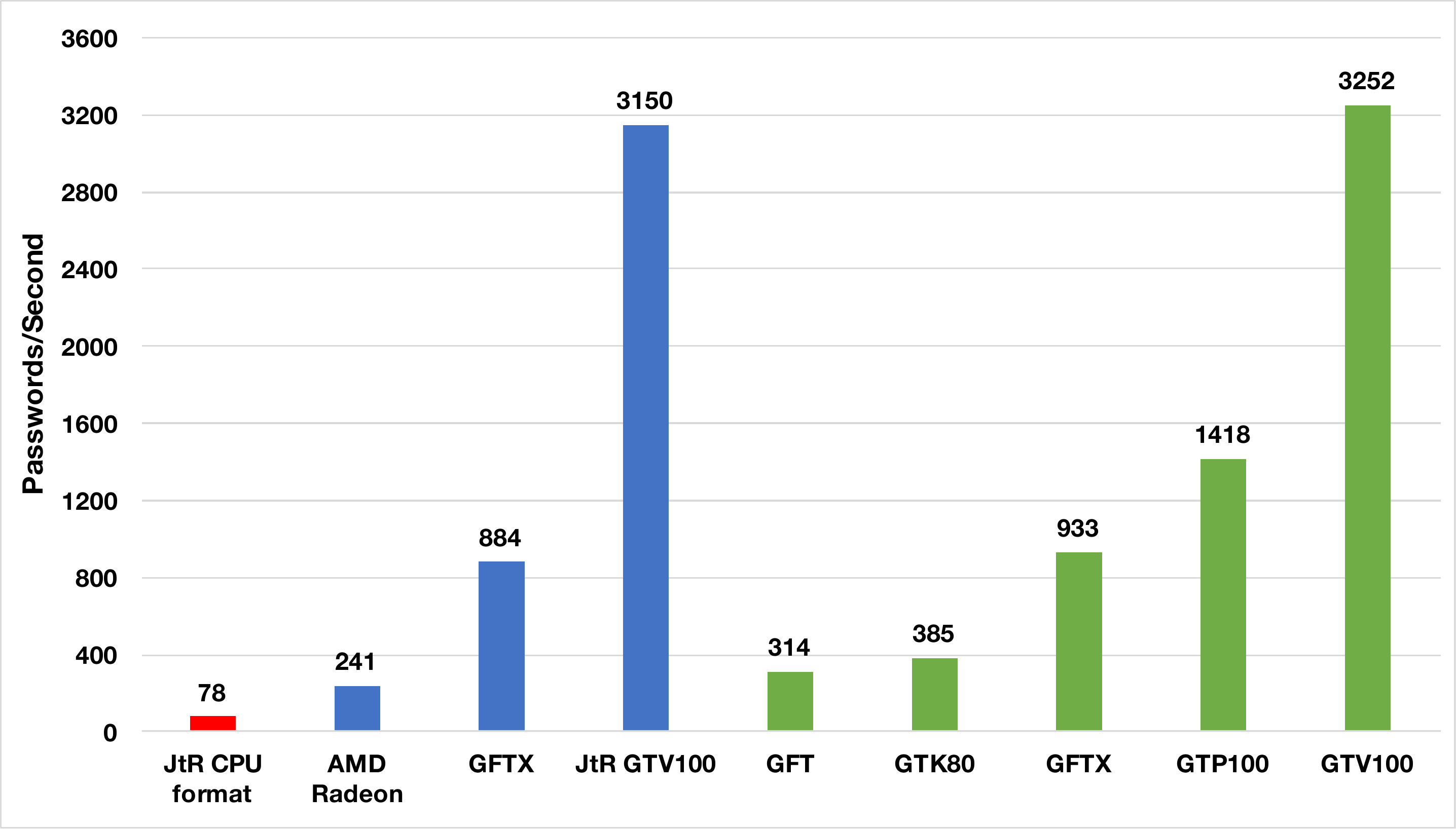}
\caption{BitCracker implementations comparison \label{fig:resume_performance}}
\end{center}
\end{figure}

In Figure \ref{fig:resume_performance} we plot the
best performance (passwords {\em per} second) obtained testing different
GPUs and software frameworks (green bars refer to the CUDA implementation whereas blue bars refer to the OpenCL implementation).

\section{Conclusions}\label{sec:conclusions}
We presented the first open-source implementation of a tool for efficient
dictionary attacks to the BitLocker crypto system.\\
The results show that our BitCracker may compete with a {\em
  state-of-the art} password cracker in terms of raw performance on
the basic computational kernels whilst it is the only one providing
specific shortcuts to speedup the BitLocker decryption procedure.
We can conclude that, although the complex architecture of BitLocker reduces
significatly the number of passwords that is possible to test in a unit of time, with
respect to other crypto-systems ({\em e.g.,} OpenPGP), it is still necessary to pay
special attention in the choice of the user password since, with a single high-end GPU,
more than a quarter-billion of passwords can be tested in a day ($\sim 3000$ passwords {\em per} second on a GTV100 $\times~86400$ seconds $\simeq 260$ million in a day).
Our implementations of SHA-256, fully customized for the CUDA-C
environment, can be reused (provided that the W words optimization is
turned off, since it cannot be applied to a general situation) for any
procedure that requires to use that hash function ({\em e.g.}, HMAC-SHA256).

Other possible improvements include the enhancement of BitCracker by
adding a mask mode attack and/or a smart reading of the input
dictionary ({\em e.g.} by assigning a probability to them) that are available
in most widely used password crackers.

We released our CUDA and OpenCL standalone implementations on GitHub
here \cite{bib:bithub}. In order to take advantage of their system of
{\em rules} for wordlist generation, our OpenCL implementation has
been released also as a John the Ripper (Jumbo version)
\cite{bib:jtr} format. We're also planning to release BitCracker within Hashcat.
%the source code can be found here \cite{bib:jtr_github} whereas the wiki reference page is here \cite{bib:jtr_wiki}.

\section{Acknowledgment}

The authors would like to thank Solar Designer and the John The Ripper team, for the access to the Openwall HPC Village.
%, and Mauro Bisson (NVIDIA) for benchmarks on the GPU Titan X with Pascal architecture.

%\appendix
%\section{SHA-256}

\end{document}